\documentclass[journal]{amsart}
\usepackage[utf8]{inputenc}
\usepackage{algorithmic,algorithm}
\usepackage{url}
\usepackage{amsmath,amssymb,amsthm}
\usepackage{graphicx}
\usepackage{fancyhdr}
\usepackage{xcolor}
\usepackage{hyperref}

\newtheorem{remark}{Remark}

\begin{document} 

\title{Time-Varying Spiky Wave-Shape Functions for Non-Stationary Signal Decomposition} 
\author{Marcelo A. Colominas and Hau-Tieng Wu}

\maketitle 

\begin{abstract}
We propose a novel framework for decomposing nonstationary spiky signals. Unlike classical adaptive non-harmonic models, which represent signals as amplitude- and frequency-modulated (AM--FM) oscillations, the proposed model is designed for signals whose dominant structures are highly localized, impulsive, or spike-like, and contains physiological variability, which is challenging to be modeled as AM--FM representations. Representative examples include electrocardiogram (ECG) complexes, epileptic electroencephalogram (EEG) transients, and other pulse-like physiological signals.
We first introduce a fixed-waveform model for repetitive spiky structures and then extend it to accommodate cycle-to-cycle waveform variability. The resulting algorithm, termed \emph{Spiky Shape-adaptive Mode Decomposition} (SSAMD), estimates waveforms in the Fourier coefficient domain. Fourier coefficients are first estimated via local regression and then regularized by exploiting their low-dimensional manifold structure. Specifically, we combine a PCA-based parametrization with an entropy regularization acting on the singular-value spectrum of the coefficient matrix, promoting morphological consistency while preserving structured waveform variability.
The proposed framework is flexible and overcomes the limitations of existing models. We validate the method on synthetic and real biomedical signals, including ECG with atrial fibrillation, epileptic EEG, and trans-abdominal maternal ECG. Experimental results demonstrate effective denoising, waveform tracking, decomposition, and segmentation, while preserving physiologically meaningful morphological variations.
\end{abstract}

\section{Introduction}

Oscillatory signal models play a central role in modern time-frequency analysis and biomedical signal processing. A widely used framework is the adaptive non-harmonic (ANH) model, in which a signal is represented as
\begin{equation}\label{eq:01}
x(t)=A(t)s(\phi(t)),
\end{equation}
where $A(t)$ is a slowly varying amplitude modulation (AM), $\phi(t)$ is the phase function with slowly varying instantaneous frequency (IF) $\phi'(t)$, and $s(t)$ is a $1$-periodic wave-shape function (WSF) \cite{wu2013instantaneous}. Since the original ANH model assumes a fixed WSF, it cannot capture cycle-to-cycle morphological variability observed in many physiological signals. This limitation has motivated extensions incorporating time-varying WSFs \cite{lin2018wave}.
Building on these models, numerous signal analysis algorithms have been developed by exploiting waveform variability. Examples include the de-shape algorithm for enhancing the fundamental component \cite{lin2018wave},  and hence decomposition mission, optimization-based mode decomposition under smoothly varying waveforms \cite{colominas2021decomposing}, and ridge extraction algorithms that explicitly account for time-varying WSFs that is robust to weak fundamental component
 \cite{su2024ridge}.
These methods have found broad applications in biomedical signal analysis, including pulse waveforms \cite{wu2016modeling}, photoplethysmography (PPG) \cite{maity2022ppgmotion}, airflow signals \cite{wu2026optimal}, etc.

Despite its success, the ANH model relies on an implicit smoothness assumption on the WSF. Specifically, the WSF is assumed to admit an efficient representation using only a small number of Fourier coefficients, or to possess rapidly decaying Fourier coefficients. This assumption is violated by many biomedical signals exhibiting highly localized transient structures, sharp peaks, or impulsive morphologies, such as ECG QRS complexes, deep-brain stimulation recordings, and cardiogenic artifacts in electroencephalography (EEG).
Although such signals can still be represented within the ANH model, doing so requires a large number of Fourier coefficients because narrow pulses exhibit slowly decaying Fourier spectra. Consequently, even modest variations in the IF produce substantial frequency variations in high-order harmonics, making many existing algorithms unstable or computationally demanding. Moreover, these signals often exhibit physiologically meaningful cycle-to-cycle waveform variability, which is difficult to capture within the ANH model framework. These limitations motivate an alternative modeling perspective that directly describes waveform evolution rather than harmonic structure.

A recent development toward this goal is the wave-shape manifold model \cite{Lin2021wsm}, which represents WSFs as points on a low-dimensional nonlinear manifold embedded in a higher-dimensional coefficient space. This geometric viewpoint has led to several manifold-based denoising and decomposition algorithms, including fetal ECG extraction from trans-abdominal recordings \cite{su2019recovery}, stimulation artifact removal from local field potential recordings during deep brain stimulation \cite{liu2026efficient}, among others. These methods are based on local manifold denoising, combining neighborhood information with optimal singular value shrinkage. More recently, the manifold framework has also been applied to change-point detection for abrupt waveform transitions \cite{colominas2023iterative} and nonparametric drift and diffusion estimation of an It\^o diffusion on the manifold by kernel regression \cite{mcerlean2026functional}.

Motivated by these developments, we study periodic signals with spiky waveforms that are difficult to analyze under the ANH model and propose a new optimization-based decomposition framework under the wave-shape manifold model. Rather than interpreting the signal through AM-FM modulation, we view it as a sequence of localized waveform events whose morphology evolves across cycles. Consequently, the primary object of interest becomes the collection of waveform shapes instead of the IF trajectory, which shifts the analysis focus from time-frequency geometry to waveform geometry.
We first introduce a fixed-waveform model capable of representing repetitive spiky events. We then extend this model by allowing the waveform coefficients to evolve on a low-dimensional manifold, thereby capturing cycle-to-cycle morphological variability directly. Unlike the ANH model, the proposed model describes waveform evolution without relying on a AM-FM interpretation.

Under this model, we develop a novel optimization-based algorithm, termed \emph{Spiky Shape-adaptive Mode Decomposition} (SSAMD). The algorithm first estimates the waveform coefficients of each cycle by linear regression and subsequently regularizes the resulting point cloud in the coefficient space. Specifically, we combine a PCA-based low-dimensional representation with an entropy regularization on the singular value spectrum of the coefficient matrix. The resulting optimization balances data fidelity with morphological consistency across cycles while preserving physiologically meaningful waveform variability.

Compared with the original ANH model \cite{wu2013instantaneous}, the proposed model accommodates cycle-to-cycle waveform variability. Compared with the time-varying WSF formulation \cite{lin2018wave}, it is specifically designed for localized, asymmetric, and spike-like waveforms that require many Fourier coefficients under the ANH model. Algorithmically, unlike existing manifold denoising methods based on local neighborhood truncation \cite{su2019recovery,liu2026efficient}, SSAMD adopts a global optimization framework in which the nonlinear manifold is approximated by a low-dimensional PCA subspace. This approximation improves robustness in the presence of nonstationary noise and artifacts while avoiding the sensitivity associated with neighborhood selection. 

We illustrate the proposed framework using both synthetic and real biomedical signals, including ECG with atrial fibrillation, EEG during epileptic events, and single-channel fetal ECG extraction problems. We compare the proposed method with existing algorithms, including the linear regression (LR) algorithm based on the ANH model \cite{wu2016modeling} and mutiresolution mode decomposition (MMD) \cite{yang2021multiresolution}. Experiments demonstrate that the proposed framework effectively performs denoising, waveform tracking, signal decomposition, and segmentation while preserving physiologically meaningful morphological variability.

\section{A Fixed-Waveform Model for Spiky Signals and its limitation} 

Periodic spiky signals arise in many applications, particularly in biomedical signal processing. Classical oscillatory models require many Fourier coefficients to properly capture their localized and structured morphology. 
We start with considering a model \cite{Lin2021wsm} with a fixed waveform:

\begin{equation}\label{model simple fixed wsf}
x(t) = \sum_i a_i \, h_T(t - t_i) \, \chi_{(t_i - T/2, t_i + T/2)}, 
\end{equation}
where $a_i\neq 0$, $t_i\in \mathbb{R}$ is a strictly increasing sequence so that $\inf_i(t_i-t_{i-1})>0$, $\chi(\cdot)$ the indicator function, $h_T(t) := h(t/T)$, and
$h(t)$ is a $1$-periodic smooth function with unit $L^2$ norm and $h(1/2)=h(-1/2)=0$. Next, model the noisy signal by a random process 
\[
x_r(t)=x(t)+\Phi(t)\,, 
\]
where $\Phi(t)$ is a white noise with mean 0 and finite variance. When $T$ is sufficiently small compared with $\min_i(t_i-t_{i-1})$, we call this model a {\em spiky waveform model with fixed waveform} (SWM-fixed). This can be viewed as a 0-dim connected wave-shape manifold \cite{Lin2021wsm}.

The goal is recovering $x(t)$ from a realization of $x_r(t)$. We solve the following optimization problem

\begin{equation} \label{equation optimization algorithm}
\begin{aligned}
&\min_{a_i,t_i,c_\ell,d_\ell,T} \bigg\| x_r(t) - \sum_{i = 1}^I a_i \chi_{(t_i - T/2, t_i + T/2)}\Big[ c_0+ \\
&   \sum_{\ell = 1}^{{D}} c_\ell \cos\Big(\frac{\ell 2 \pi (t-t_i)}{T}\Big) + d_\ell \sin \Big(\frac{\ell 2 \pi (t-t_i)}{T}\Big) \Big] \bigg\|_2^2, 
\end{aligned}
\end{equation}
where we approximate the waveform $h(t)$ using the first ${D}$ harmonics. Notice that all the parameters are real-valued, even when working on a discrete-time setting. 

\begin{remark}
We shall comment that under the traditional ANH model, a spiky WSF $h_T$ needs a large $D$, depending on how spiky $h_T$ is. In our new formula, the approximation is limited locally to the scale of $T$, so we may not need a large $D$ in the trade of estimating an extra parameter $T$. 
\end{remark}

\subsection{initialization for solving \eqref{equation optimization algorithm}} 

The non-linear regression problem \eqref{equation optimization algorithm} needs initial values $\{a_i^0\}_{i = 1}^I$, $\{t_i^0\}_{i = 1}^I$, $\{c_\ell^0\}_{\ell = 0}^D$, $\{d_\ell^0\}_{\ell = 1}^D$, and $T^0$. Good initial estimations lead to a warm-start strategy \cite{yildirim2002warm} that produces satisfactory results. We call this algorithm {\em Initialization of the optimization} (InitOpt). Also, the number of cycles $I$ needs to be determined, as well as the number of harmonics $D$. 

\subsubsection{Initial estimations} 
In this step, we model the signal using the ANH model  \eqref{eq:01} and run linear regression \cite{wu2013instantaneous,wu2016modeling} to obtain a prior estimate. Write  $x(t)\approx A(t) s(\phi(t))$, where $s(t) = \sum_{\ell = 0}^D \alpha_\ell \cos(\ell 2 \pi t) + \beta_\ell \sin(\ell 2\pi t)$ for some $D\in \mathbb{N}$. Under this model, $A(t)$ and $\phi(t)$ can be well estimated by applying the second-order synchrosqueezing transform \cite{daubechies2011synchrosqueezed,oberlin2015second}, $D$ can be estimated by applying trigonometric regression \cite{ruiz2022wave}, obtaining $\hat{D}$. This prior estimate is in general limited under the current spiky model but offers useful information (see Results). From the estimated phase $\phi(t)$, we obtain the number of cycles $\hat{I}$.

\subsubsection{Initial scale $T^0$} We use the inverse of the average of the estimated $\phi'(t)$.

\subsubsection{Initial central time instants $\{t_i^0\}_{i = 1}^{\hat{I}}$} 

It is in general challenging to obtain an accurate estimation of $\{t_i^0\}_{i = 1}^{\hat{I}}$ from $\phi(t)$. In fact, if $t_i^0$ satisfying $\phi(t_i^0)=i+\phi_0$ for some prescribed global phase $\phi_0\in [0,1)$ is an estimator of $t_i^0$, it is usually biased since the precise location also depends on high order harmonics but we only obtain low order harmonics. See \cite{Lin2021wsm,Alian2022reconsider} for some discussion. Our solution is applying the \emph{matching pursuit} approach \cite{mallat1993matching}. Knowing the number of cycles ${\hat{I}}$ and using the rough estimation $s(t)$, a matching pursuit strategy retrieves good estimations of $\{t_i^0\}_{i = 1}^{\hat{I}}$. 

\subsubsection{Initial amplitudes $\{a_i^0\}_{i = 1}^{\hat{I}}$} With the central time instants and the estimation of the amplitude modulation, we can estimate the initial amplitudes as $\{a_i^0\}_{i = 1}^{\hat{I}} = \{A(t_i^0)\}_{i = 1}^{\hat{I}}$. 

With the above initial values and parameters, run the optimization \eqref{equation optimization algorithm}. We call the composition of InitOpt and the solving of \eqref{equation optimization algorithm}  `SWM-fixed-Alg'. Note that InitOpt is the initialization of the main proposed algorithm that will be detailed in the next section.

\begin{remark}
The approximation of \eqref{model simple fixed wsf} by \eqref{eq:01} in the {\em initial estimations} is in general rough. Note that we do not assume any regularity of $t_i-t_{i-1}$, which means $\phi$ can vary fast. For example, when the subject has atrial fibrillation, $t_i-t_{i-1}$ estimated from the ECG behaves like a shifted white noise \cite{hennig2006exponential}. In this case, the estimated $\phi$ by SST provides a ``smoothened'' version of $\phi$, which is inaccurate but useful for us to obtain $I$ and a prior WSF estimate. When the WSF $h_T$ is spiky, $D$ in general can be infinite. In this case, the trigonometric regression leads to an underestimated $D$. Again, this prior estimate is inaccurate but useful, particularly for the central time instants determination.
\end{remark}

\subsection{Simulations and Results for the Fixed-Waveform Model} 

We provide here some  results. We test our proposal on  synthetic signals. We use a train of Gaussian functions, and a train of Gaussian derivatives, and a train of a synthetic ECG cycle. For all three cases, each cycle is \emph{not} a modulated waveform, but the exact same waveform without any stretching.

\begin{figure}[ht!]
\begin{center}
\includegraphics[width=\columnwidth]{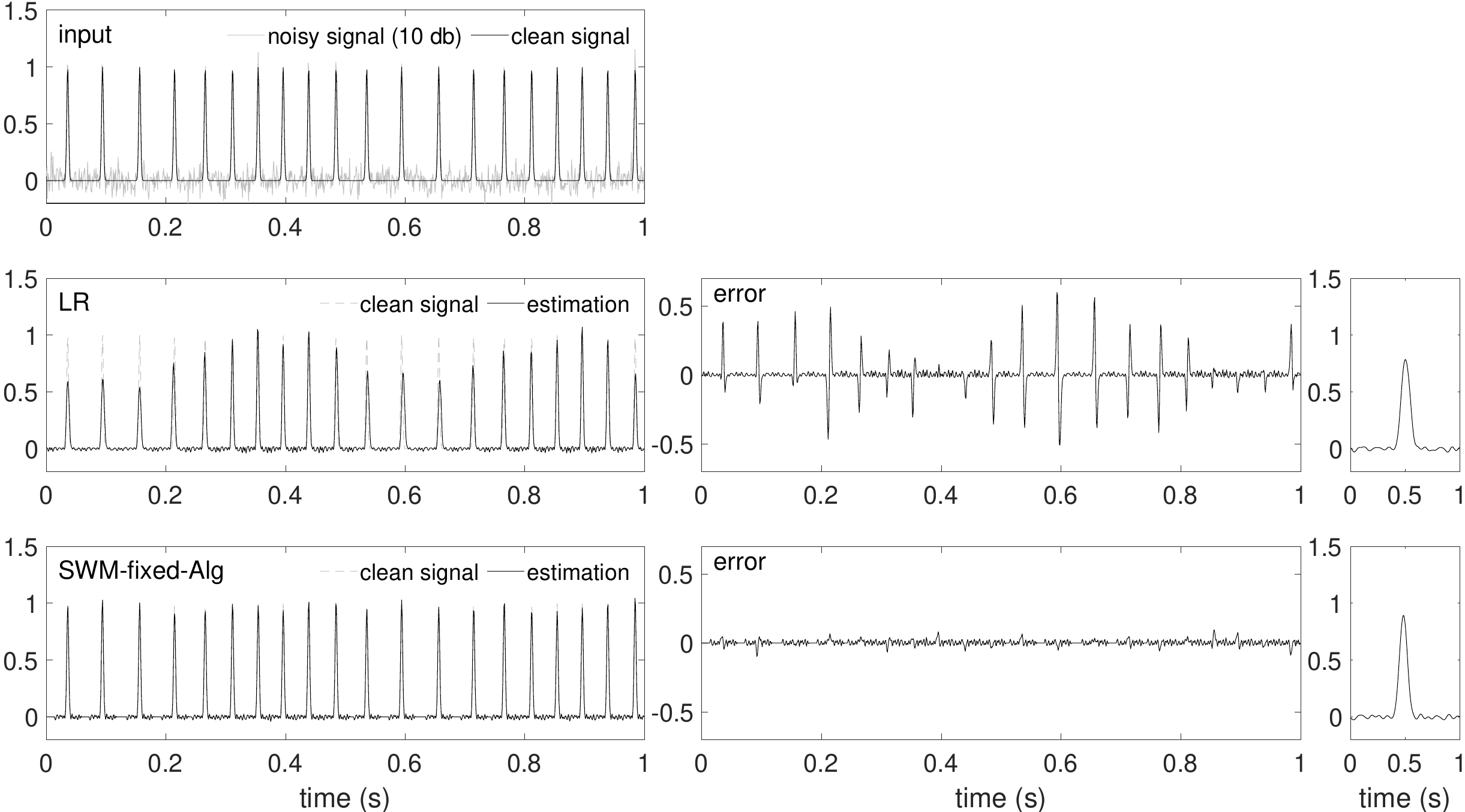} 
\end{center} 
\caption{\textbf{Train of Gaussian functions.} Top Row: input noisy signal (10 db), and clean signal as a reference. Middle Row: LR results, its error, and its estimated waveform. Bottom Row: SWM-fixed-Alg results, its error, and its estimated waveform.} \label{fig:01} 
\end{figure}

\textbf{Train of Gaussian functions.} Here our signal is $x(t) = \sum_{i = 1}^{20} e^{-20000(t-t_i)^2} + n(t)$, for $0 \leq t \leq 1$, with $n(t)$ a white Gaussian noise with amplitude such that the SNR is 10 dB. Here, the time instants $t_i$ are such that $\phi(t_i) = k 2 \pi$ with $k = 1,\dots,20$ and $\phi(t) = 2\pi 20 t + 2 \cos(4 \pi t)$. Notice that the points $t_i \in \mathbb{R}$ are \emph{off} the sampling grid, and as such it seems as there is a small amplitude modulation (but there is not, it is only a sampling effect). The results are presented on Fig. \ref{fig:01}. We show the clean signal, and its noisy version in the top row, results with the ANH model, and results with the new model. We also show the errors (between the estimations and the clean signal) and the estimated waveforms.  Although these last two might seem similar, the new model estimated waveform is actually narrower, being closer to the clean waveform used to synthesize the signal.

\begin{figure}[] 
\begin{center} 
\includegraphics[width=\columnwidth]{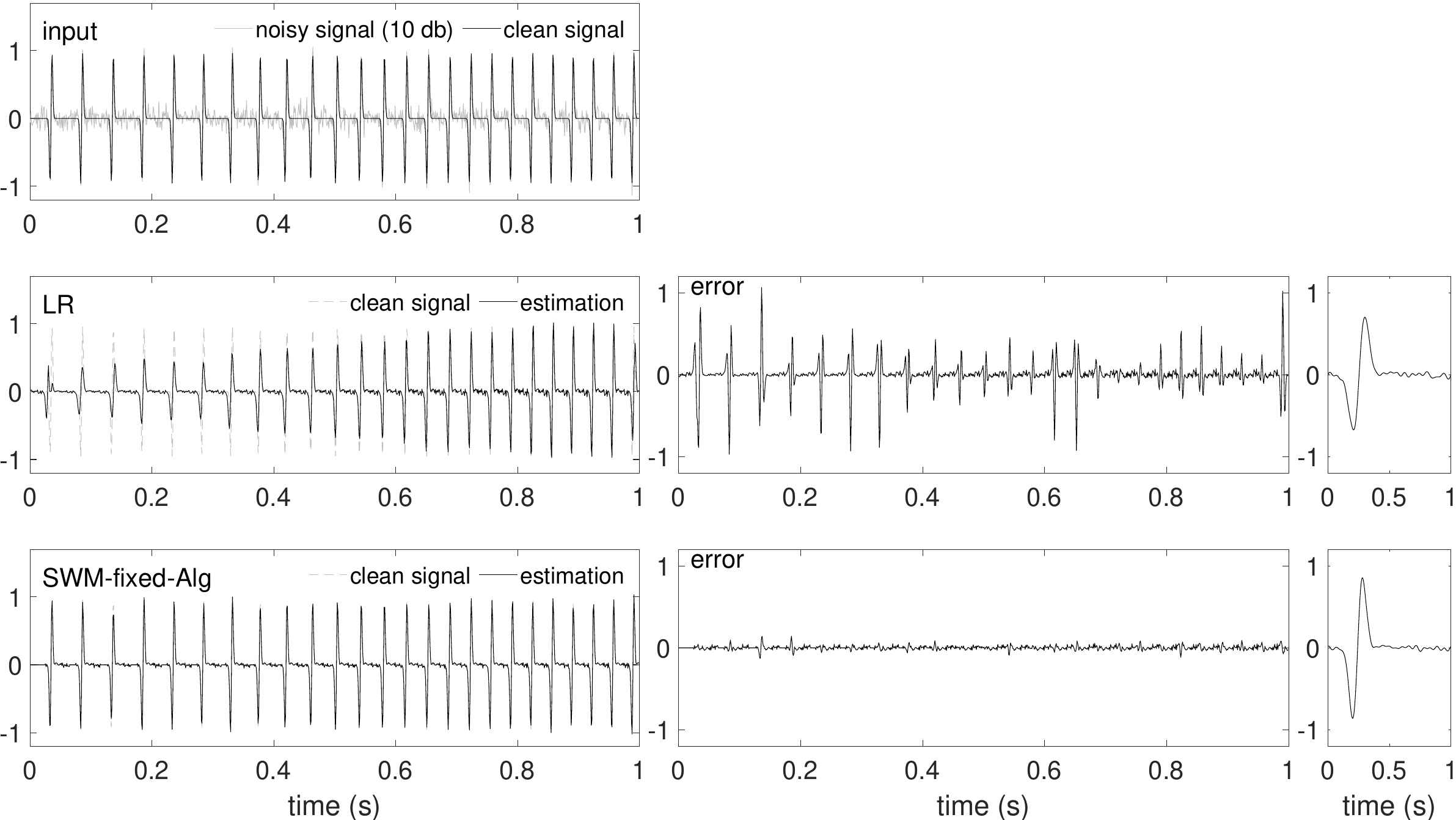} 
\end{center} 
\caption{\textbf{Train of Gaussian derivatives.} Top Row: input noisy signal (10 db), and clean signal as a reference. Middle Row: LR results, its error, and its estimated waveform. Bottom Row: SWM-fixed-Alg results, its error, and its estimated waveform.} \label{fig:02} 
\end{figure}

\textbf{Train of Gaussian derivatives.} For this example our signal is $x(t) = \sum_{i = 1}^{25} 1000(t-t_i) e^{-20000(t-t_i)^2} + n(t)$, for $0 \leq t \leq 1$, with $n(t)$ a white Gaussian noise with amplitude such that the SNR is 10 dB. Here, the time instants $t_i$ are such that $\phi(t_i) = k 2 \pi$ with $k = 1,\dots,25$, and $\phi(t) = 2\pi 20 t + 2 \cos(2 \pi t) + 2\pi 5 t^2$. In this case, the frequency increases with time, and that makes the waveforms closer to each other as time passes. Notice that, as before, the points $t_i \in \mathbb{R}$ are \emph{off} the sampling grid, and as such it seems as there is a small amplitude modulation (but there is not, it is only a sampling effect). The results are presented on Fig. \ref{fig:02}. We show the clean signal, and its noisy version. We show the LR results, and the new model. We also show the errors (between the estimations and the clean signal) and the estimated waveforms. 

\begin{figure}[h!] 
\begin{center} 
\includegraphics[width=\columnwidth]{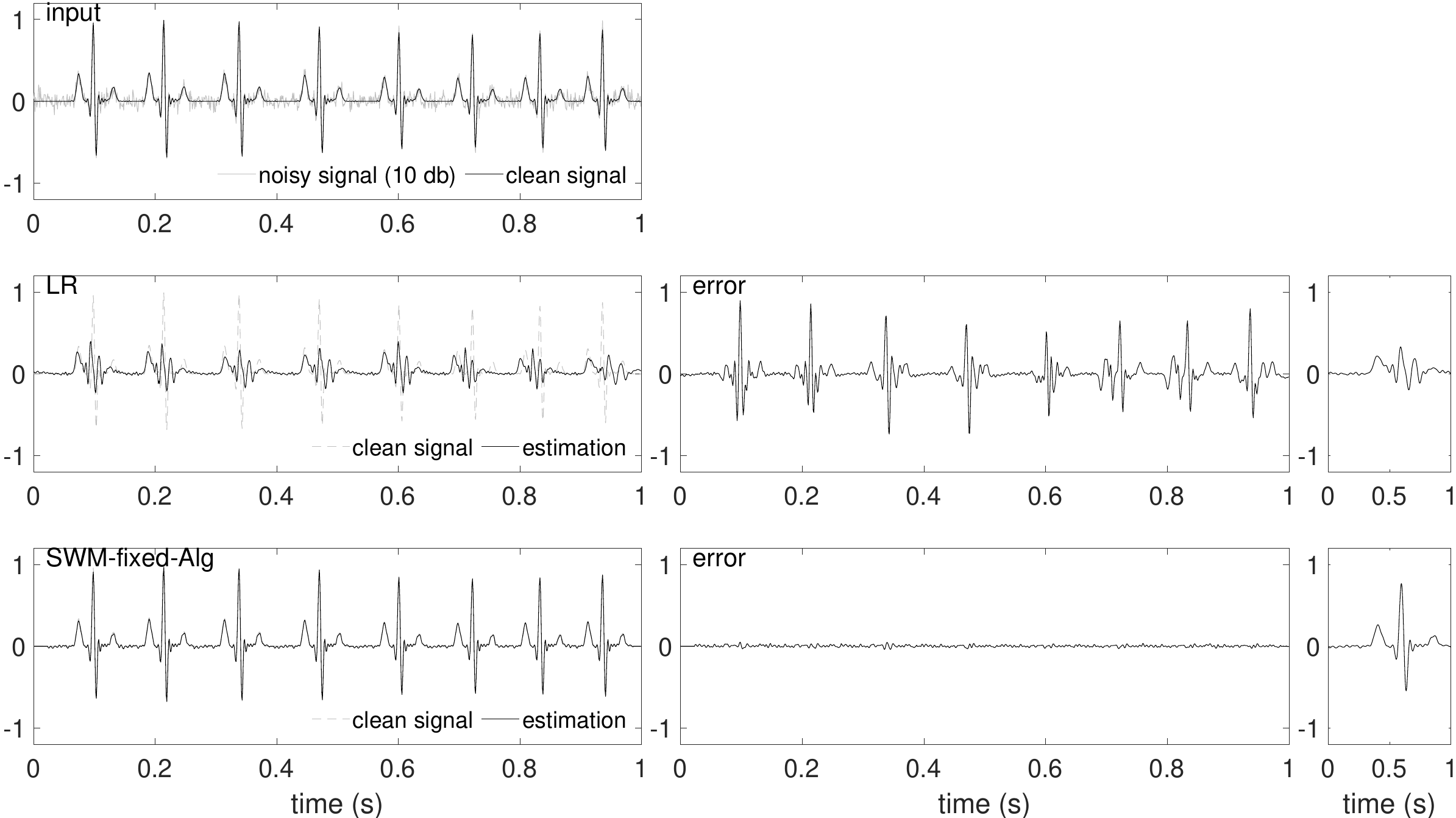} 
\end{center} 
\caption{\textbf{Train of synthetic ECG cycles.} Top Row: input noisy signal (10 db), and clean signal as a reference. Middle Row: LR results, its error, and its estimated waveform. Bottom Row: SWM-fixed-Alg results, its error, and its estimated waveform.} \label{fig:03} 
\end{figure}

\textbf{Train of synthetic ECG cycles.} Here our signal is $x(t) = \sum_{i = 1}^8 a_i h(t-t_i) + n(t)$, for $0 \leq t \leq 1$, with $n(t)$ a white Gaussian noise with amplitude such that the SNR is 10 dB. Here, the function $h(t)$ is a synthetic ECG cycle. The time instants $t_i$ are \emph{on} the sampling grid, and are such that $\phi(\tau_i) = k 2 \pi$, with $k = 1,\dots,8$, $t_i$ is the projection of $\tau_i$ onto the grid, and $\phi(t) = 2\pi 8 t + \sin(2 \pi t) + \pi t^2$. We incorporated an amplitude modulation such that $a_i = 1 + 0.1 \sin(2\pi t_i)$. The results are presented on Fig. \ref{fig:03}. We show the clean signal, and its noisy version. We show the LR results, and the new model. We also show the errors (between the estimations and the clean signal) and the estimated waveforms. 

\section{A General Model for Time-Varying Spiky Waveforms}

In many spiky signals (e.g., biomedical or physiological recordings), individual cycles exhibit morphological changes that cannot be captured by a single template. See \cite{lin2018wave} for a discussion. We now extend the SWM with fixed-waveform to accommodate waveform variability across cycles. Consider the following model:
\begin{equation}
y(t) = \sum_{i=1}^{I} a_i h_i(t - t_i)\,\chi(t_i-T/2,\, t_i+T/2),
\label{eq:model_tv}
\end{equation}
where $a_i, t_i$, $\chi(\cdot)$ are the same as those in \eqref{model simple fixed wsf}, while $h_i(t)$ are time-dependent waveforms that is assumed to be smooth $T$-periodic with truncated Fourier representation
\begin{equation}
h_i(t) = c_{0,i} + \sum_{\ell = 1}^D c_{\ell,i} \cos(\ell 2\pi t/T) + d_{\ell,i} \sin(\ell 2\pi t/T)
\end{equation}
and $h_i(-T/2)=h_i(T/2)=0$. We assume $\{h_i\}$ can be parametrized by a low dimensional smooth wave-shape manifold embedded in the Hilbert space $L^2([-T/2,T/2])$ \cite{Lin2021wsm}. Take a random process 
\begin{equation}\label{Model noisy time-varying WSF model}
y_r(t)=y(t)+\Phi(t)\,, 
\end{equation}
where $\Phi(t)$ is a white noise with mean 0 and finite variance. When $T$ is sufficiently small compared with $\inf_i(t_i-t_{i-1})>0$, we call this model a {\em spiky waveform model with time-varying waveform} (SWM-variable).

\section{Proposed Algorithm: Spiky Shape-Adaptive Mode Decomposition (SSAMD)}

Our goal is to recover $y(t)$ from a realization of $y_r(t)$ \eqref{Model noisy time-varying WSF model}. The proposed algorithm is composed of two steps. In Step 1, we apply the same optimization algorithm in \eqref{equation optimization algorithm}. Note that most parameters, denoted as $\{\hat a_i\}_{i = 1}^{\hat I}$, $\{\hat t_i\}_{i = 1}^{\hat I}$, and $\hat T$, are output of \eqref{equation optimization algorithm}. However, due to the time-varying nature of WSFs, the estimation of Fourier coefficients, $\{\hat c_{\ell},\hat d_{\ell}\}$, in Step 1 are biased. 
In Step 2, after running the optimization algorithm in \eqref{equation optimization algorithm}, we apply a singular value decomposition (SVD) entropy regularization to refine the coefficients  $\{c_{\ell},d_{\ell}\}$ and 
recover time-varying waveforms obtaining $\{c_{\ell,i},d_{\ell,i}\}$.
The pseudocode of the second step is summarized in Algorithm \ref{alg:svd_entropy}. We call the algorithm composing of steps 1 and 2 {\em spiky shape-adaptive mode decomposition} (SSAMD). 
We now detail Step 2.

\subsubsection{Initial step}
Construct
\begin{equation}
X_d := [\mathbf{c}_1^d,\mathbf{c}_2^d,\dots,\mathbf{c}_{\hat{I}}^d] \in \mathbb{R}^{(2\hat{D}+1)\times {\hat{I}}},
\end{equation}
where 
\begin{equation}
    \mathbf{c}_i^d = \arg \min_{\mathbf{c}\in \mathbb{R}^{(2\hat{D}+1)\times 1}} \|x_{r,i}(t) - \mathbf{Q}_i \mathbf{c}  \|^2 \in \mathbb{R}^{(2\hat{D}+1)\times 1}\,,
\end{equation}
$x_{r,i}(t) := x_r(t-\hat t_i)$ with $t\in [-\hat T/2, \hat T/2]$, and
\begin{equation}
\mathbf{Q}_i := [\{a_i \cos(2\pi t/\hat T)\}_{i=1}^{\hat I}, \{a_i \sin(2\pi t/\hat T)\}_{i=1}^{\hat I}]
\end{equation}
contains $2\hat{D}+1$ functions on $[-\hat T/2,\hat T/2]$ on the columns.
Note that this procedure is a local version of the {\em initial estimations} of Step 1 in the sense that we estimate the Fourier coefficients cycle-by-cycle. It is merely a \emph{linear regression on each cycle} and would include artifacts and noise. 

Next, use $X_d$ as a reference and leverage the low-dimension structure favoring homogeneity between the cycles to recover $x(t)$. 
Denote
\begin{equation}
X_0 = \bar{\mathbf{c}}\mathbf{1}^\top\in \mathbb{R}^{(2\hat D+1)\times \hat I},
\end{equation}
where $\mathbf{1}\in \mathbb{R}^{\hat I}$ is a vector with all entries $1$, $\bar{\mathbf{c}}\in \mathbb{R}^{(2\hat D+1)\times 1}$ is the coefficient spectrum determined by the fixed-waveform model, i.e. coefficients $\{c_\ell\}_{\ell = 0}^{\hat{D}}$ and $\{d_\ell \}_{\ell = 1}^{\hat{D}}$ of the output of \eqref{equation optimization algorithm}. We design a functional depending on $X_d$, and recover $x(t)$ by optimization. 

\subsubsection{Objective functional}
Let $V \in \mathbb{R}^{(2\hat D+1)\times 3}$ be an orthonormal basis obtained from the first three principal components of $X_d - X_0$. We parametrize our desired solution as 
\begin{equation}\label{model manifold constraints 3dim}
X:=X(A) = X_0 + VA  , \qquad A \in \mathbb{R}^{3\times \hat I}.
\end{equation}
This guarantees
\begin{equation}  
\mathrm{rank}(X - X_0) \le 3 .
\end{equation}
We define the objective functional $\Phi:\mathbb{R}^{3\times \hat I}\to \mathbb{R}_+$ by
\begin{equation}
\Phi(A) = \Phi_{\text{rep}}(A) + \lambda \Phi_{\text{hom}}(A)\,,
\end{equation}
where $\Phi_{\text{rep}}(A):= \| X(A) - X_d \|_F^2$ is called {\em representation error term}, which represents how faithful to the input data the output will be, and
$\Phi_{\text{hom}}(A)$ is called the {\em homogeneity term}, which is defined in the following way. Let the singular value decomposition (SVD) of the centered matrix be
\[
X(A) - X_0 = U \Sigma W^\top ,
\]
where the singular values are denoted as $\sigma_1\geq \sigma_2\geq \ldots$. Define the normalized spectrum as
\begin{equation}
p_i = \frac{\sigma^2_i}{\sum_j \sigma^2_j}
\end{equation}
the {\em SVD entropy} \cite{alter2000singular} as
\begin{equation}
H(A) = -\sum_i p_i \log p_i ,
\end{equation}
and the {\em effective dimension} ass  $d_{\text{eff}}(A) := e^{H(A)}$ \cite{cover2006elements,roy2007effective}.
Set
\begin{equation}
\Phi_{\text{hom}}(A) = H(A)\,,
\end{equation}
which acts as a regularization term to prevent the output to merely mimic the input data. Notice that $H(X-X_0) =  H(VA) = H(A)$.

\subsubsection{{Optimization problem}} Our optimization problem will be the following:
\begin{equation}
A^* = \arg\min_A \left( \| X_0 + V A - X_d \|_F^2 + \lambda H(A) \right)\,,
\end{equation}
where
the parameter $\lambda\geq 0$ controls the homogeneity of the point cloud within the fixed three-dimensional subspace. Small $\lambda$ yields configurations close to the data with $d_{\text{eff}}\approx 3$, while large $\lambda$ produces spectrally concentrated configurations with $d_{\text{eff}}\approx 1$.
We consider the following automatic selection of $\lambda$.
Given a target effective dimension $d^* \le 3$, set 
\begin{equation}\label{eq:lambda}
\lambda^* = \arg\min_\lambda | d_{\text{eff}}(\lambda) - d^* |.
\end{equation}

\subsubsection{{Gradient Descent}} We will solve the optimization problem \emph{via} gradient descent.
The gradient flow reads:
\begin{equation}
A^{k+1}=A^k-\eta\nabla_A \Phi(A^k),
\end{equation}
for the $(k+1)$-th iteration, where $\eta>0$ is the learning rate and
\begin{equation}
\nabla_A \Phi = 2 V^\top (X-X_d) + \lambda\nabla_A\Phi_{\text{hom}}.
\end{equation}
The derivation of $\nabla_A \Phi_{\text{hom}}$ is postponed to Appendix.
The algorithm is run for several values of $\lambda$, and the final solution is the one that satisfies \eqref{eq:lambda}. 
In brief, we start from $X_0$ and move \emph{towards} $X_d$ without reaching it. 
As we will show below, this procedure is flexible to empirically accommodate the cycle-to-cycle variations of real biomedical signals.
With $A^*$, the final time-varying WSFs are estimated by
\begin{equation}
X^*:=X_0 + VA^*=[c^*_1,\ldots,c^*_{\hat{I}}] \,. 
\end{equation}
The $i$-th WSF is estimated by $\hat{x}_{\hat{T},i}:=\mathbf{Q}_i c^*_i$, and hence the recovered $x(t)$.

To summarize, we reduce the problem of estimating the time-varying spiky WSF to ``regularizing'' or denoising a noisy point cloud in the Fourier coefficients space. Essentially, SSAMD is a gradient descent based optimization solver with SVD-entropy regularization. 

\subsection{Comparison with existing algorithms}

$X_d$ is a noisy random matrix, and recovering the time-varying WSFs can be viewed as a matrix denoising problem. A classical approach is based on the SVD,
\begin{equation}
X_d = U_d \Sigma_d V_d^\top,
\end{equation}
followed by a rank-$r$ approximation
\begin{equation}
X_{d,r} = U_{d,r} \tilde{\Sigma}_{d,r} V_{d,r}^\top\,,
\end{equation}
where $\tilde{\Sigma}_{d,r}$ is obtained by either hard-thresholding the singular values \cite{eckart1936approximation} or applying an optimal shrinkage \cite{gavish2017optimal,su2025data}.
The rank-1 approximation is useful when the WSF is fixed, and therefore recovers the fixed-waveform model. Increasing the rank allows more morphological variability. In practice, waveform morphology often evolves in a nonlinear manner. This observation motivates modeling the WSFs as lying on a low-dimensional nonlinear manifold $\mathcal{M}$ embedded in $\mathbb{R}^{2D+1}$ rather than a linear subspace \cite{hastie1989principal,gorban2008principal}. Based on this perspective, manifold-based decomposition and denoising algorithms have been developed \cite{su2019recovery,liu2026efficient}. These methods exploit the local geometry of $\mathcal{M}$ and denoise each noisy sample using only its neighboring points on the manifold, instead of the entire dataset. We refer interested readers to these papers and the references therein for further details. 
Compared with SVD-based linear methods, manifold-based approaches provide greater modeling flexibility and are particularly well suited to physiological signals whose waveform morphology evolves continuously from cycle to cycle.

Note that, for the truncation-based manifold denoising algorithm in \cite{su2019recovery,liu2026efficient}, the {\em effective rank} determined by the optimal shrinkage procedure plays a role analogous to the effective dimension $d_{\text{eff}}(A)$, which quantifies the local morphological complexity of the data. See \cite[Table 1]{10.1214/17-AOS1601} for a relevant discussion. Unlike these methods, which fully exploit the nonlinear manifold structure, the proposed SSAMD constraints the nonlinear manifold to a 3-dim linear subspace through the projection in \eqref{model manifold constraints 3dim}. This additional constraint provides a favorable bias-variance tradeoff by reducing estimation variance at the cost of a modest increase in modeling bias. Empirically, we find that this constraint improves robustness in the presence of nonstationary noise and artifacts. We also observe that a $3$-dim linear subspace is sufficient for a wide range of real-world datasets, although the subspace dimension can be adjusted when warranted by a particular application or when the variability of WSFs is large. 

\begin{algorithm}
\caption{Gradient Descent with SVD-Entropy Regularization}
\label{alg:svd_entropy}
\begin{algorithmic}[1]
\REQUIRE Reference matrix $X_d\in\mathbb{R}^{(2\hat D+1)\times \hat I}$, initial matrix $X_0=\bar{\mathbf{c}}\mathbf{1}^\top$, PCA basis $V\in\mathbb{R}^{(2\hat D+1)\times 3}$, regularization parameter $\lambda>0$, step size $\eta>0$, maximum iterations $N_{\max}$, tolerance $\varepsilon$
\ENSURE Estimated coefficient matrix $X$

\STATE Initialize $A^{(0)} = 0 \in \mathbb{R}^{3\times \hat I}$
\STATE $k \gets 0$

\WHILE{$k < N_{\max}$}

    \STATE Compute current estimate:
    \[
    X^{(k)} = X_0 + V A^{(k)}
    \]

    \STATE Compute SVD:
    \[
    VA^{(k)} = U^{(k)} \Sigma^{(k)} (W^{(k)})^\top
    \]

    \STATE Compute probabilities:
    \[
    p_i^{(k)} =
    \frac{(\sigma_i^{(k)})^2}
    {\sum_j (\sigma_j^{(k)})^2}
    \]

    \STATE Compute entropy:
    \[
    H^{(k)} =
    -\sum_i p_i^{(k)} \log p_i^{(k)}
    \]

    \STATE Compute gradient of entropy term:
    \[
    G_H^{(k)} =
    U^{(k)}
    \mathrm{diag}\!\left(
    -\frac{2\sigma_i^{(k)}}{S^{(k)}}
    (\log p_i^{(k)} + H^{(k)})
    \right)
    (W^{(k)})^\top
    \]
    where
    \[
    S^{(k)}=\sum_j (\sigma_j^{(k)})^2
    \]

    \STATE Compute total gradient:
    \[
    G^{(k)} =
    2V^\top (X^{(k)} - X_d)
    + \lambda G_H^{(k)}
    \]

    \STATE Gradient update:
    \[
    A^{(k+1)} =
    A^{(k)} - \eta G^{(k)}
    \]

    \IF{$\|A^{(k+1)} - A^{(k)}\|_F < \varepsilon$}
        \STATE \textbf{break}
    \ENDIF

    \STATE $k \gets k+1$

\ENDWHILE

\STATE \RETURN $X^{(k)} = X_0 + V A^{(k)}$

\end{algorithmic}
\end{algorithm}


\section{Numerical Results}

We present some examples on both simulated and real signals. We will test algorithms developed on the original ANH model, SWM-fixed model, and SWM-variable model. The Matlab code can be found in \url{https://github.com/macolominas/Spiky-SAMD}.

In all the examples we used a step size $\eta = 0.1$, a target dimension $d^* = 2$, and a maximum number of iterations $N_{\max} = 100$. For the regularization parameter $\lambda$ we explored a range from $10^{-4}$ to $10^2$ in $100$ steps in logarithmic scale. We also compare with multiresolution mode decomposition (MMD) \cite{yang2021multiresolution}, which is developed upon an oscillatory model that takes waveform variability into account as well.

\subsection{Real EEG signal during epileptic event}

\begin{figure}[h!] 
\begin{center}
\includegraphics[width=\columnwidth]{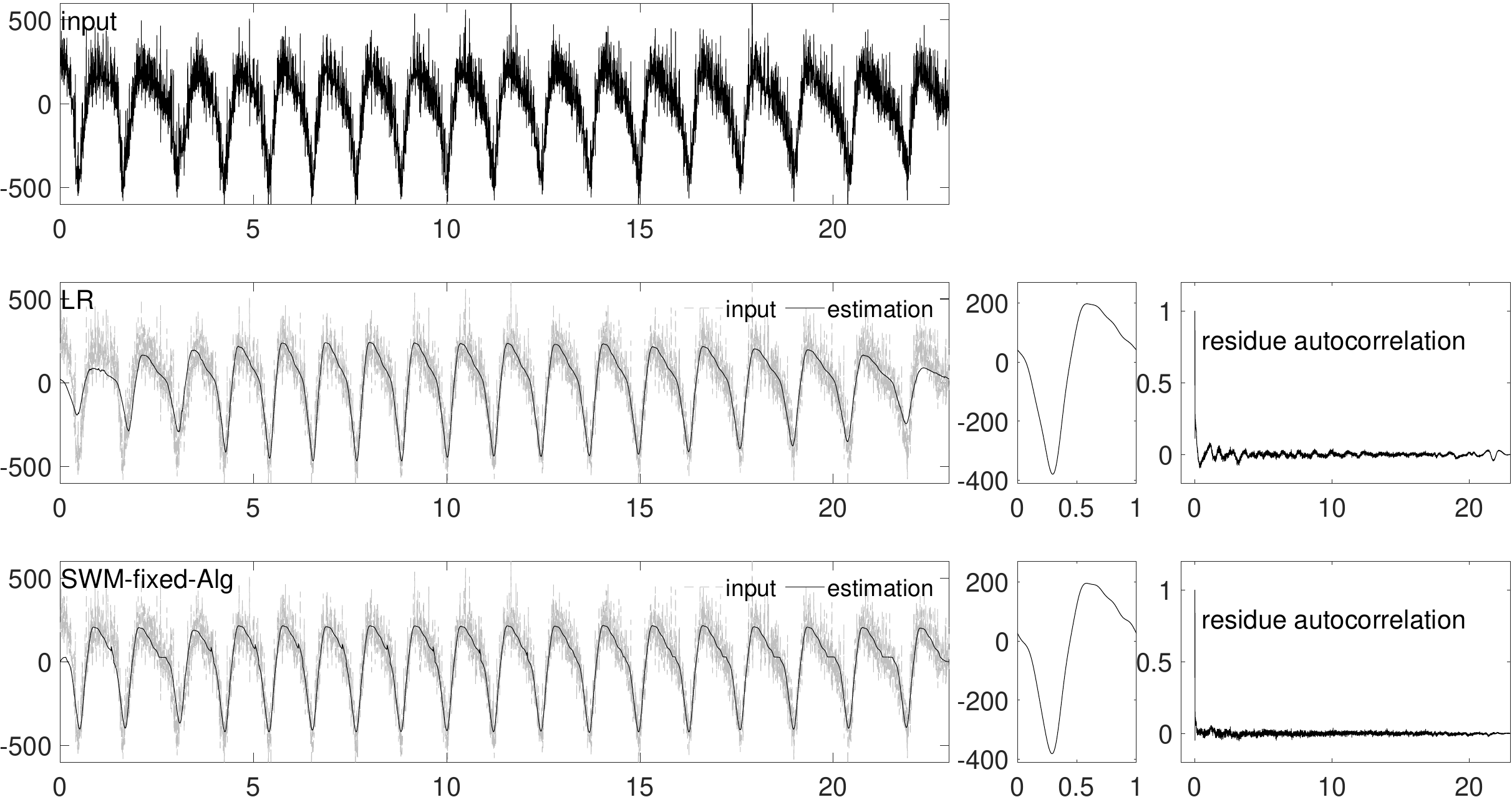}

\vspace{2.5mm}
\includegraphics[width=\columnwidth]{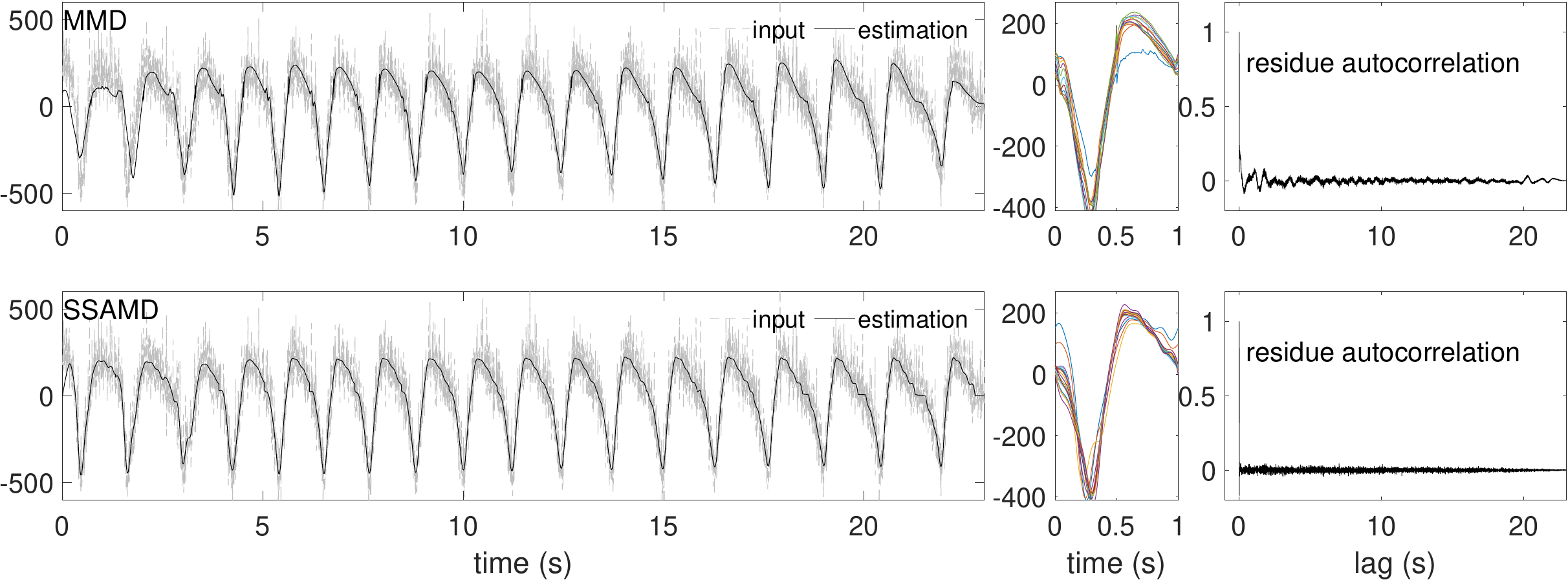}
\end{center} 
\caption{\textbf{EEG Recording from \cite{stevenson2019dataset}.} Top row: input signal. Second row: LR results, its estimated waveform, and the residue autocorrelation. Third row: SWM-fixed-Alg results, its estimated waveform, and the residue autocorrelation. Fourth row: MMD results, its estimated waveforms, and the residue autocorrelation. Fifth row: SSAMD results, its estimated waveforms, and the residue autocorrelation.} \label{fig:EEG} 
\end{figure}

In this example, we analyze an electroencephalography (EEG) signal recorded from a newborn during an epileptic seizure episode, obtained from a publicly available dataset \cite{stevenson2019dataset}. The EEG signal and the results of different algorithms are shown in Fig. \ref{fig:EEG}. A strong presence of ictal rhythmic activity, associated with the seizure, can be observed superimposed with noise originating from muscular contraction \cite{stevenson2019dataset}. For LR, only a single waveform can be estimated. We can also see that, while offering good results in general, it tends to fail near the borders of the signal due to the lack of flexibility of handling different WSFs. Our new SWM-fixed-Alg is able to obtain better results near the boundary, but the best performance is achieved by SSAMD (since this allows to accommodate variable waveforms). In short, SSAMD is capable of capturing time-varying waveforms, leading to a better denoising performance. 

On the right column of Fig. \ref{fig:EEG}, we show the autocorrelation functions of the residues (difference between the input and the estimation), to evaluate how ``white'' the residues are. The result of SSAMD is superior, while that of SWM-fixed-Alg is still acceptable.  The MMD method produces a noisier estimation whose residue is more correlated.
Numerically, for this 23 s long signal sampled at 256 Hz, SWM-fixed-Alg takes 6.876 s, and SSAMD takes 7.251s, while MMD take 11.170 s.

\subsection{Real ECG signals} 
\begin{figure}[h!] 
\begin{center} 
\includegraphics[width=\columnwidth]{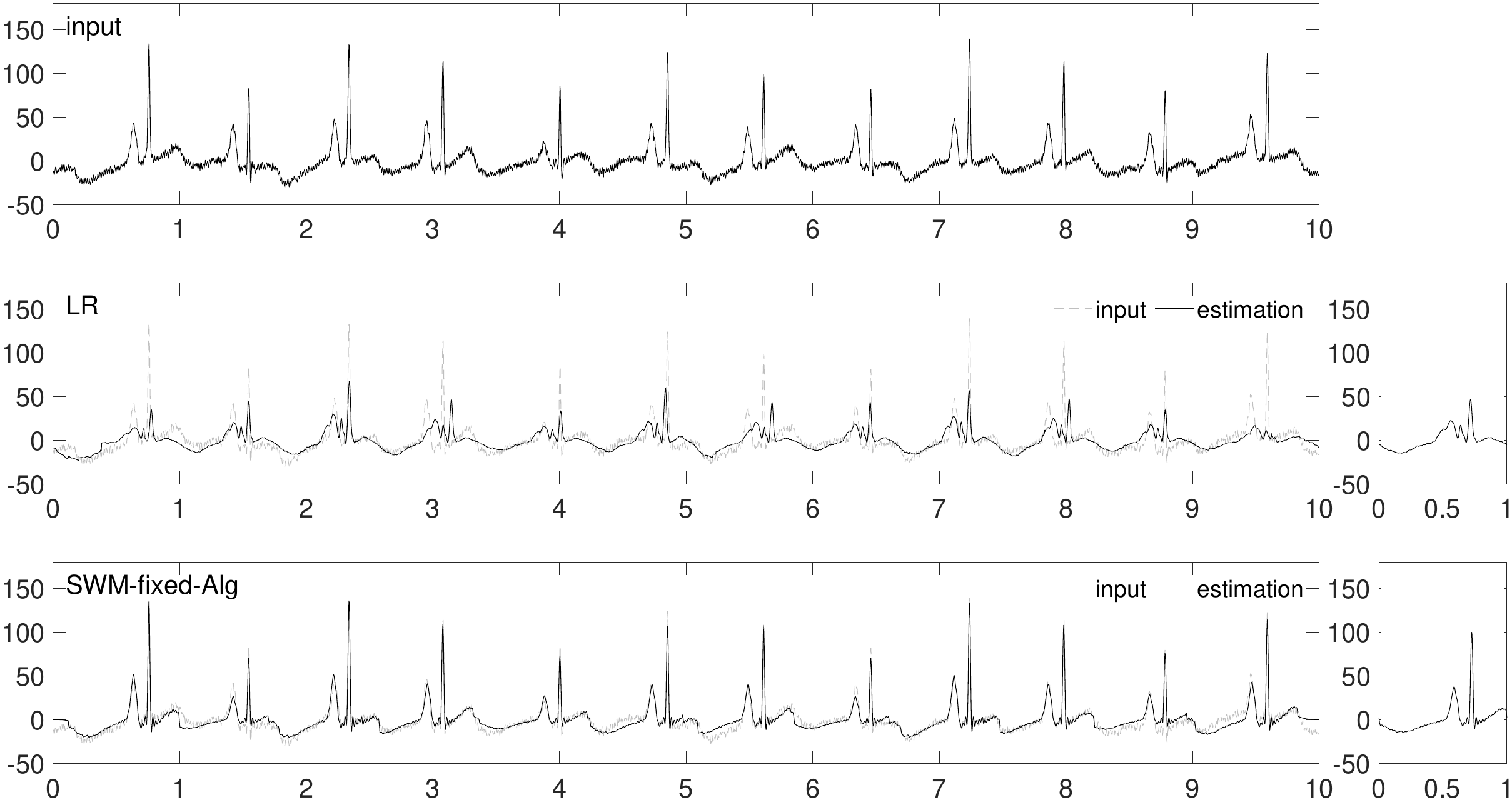}

\vspace{2.5mm}
\includegraphics[width=\columnwidth]
{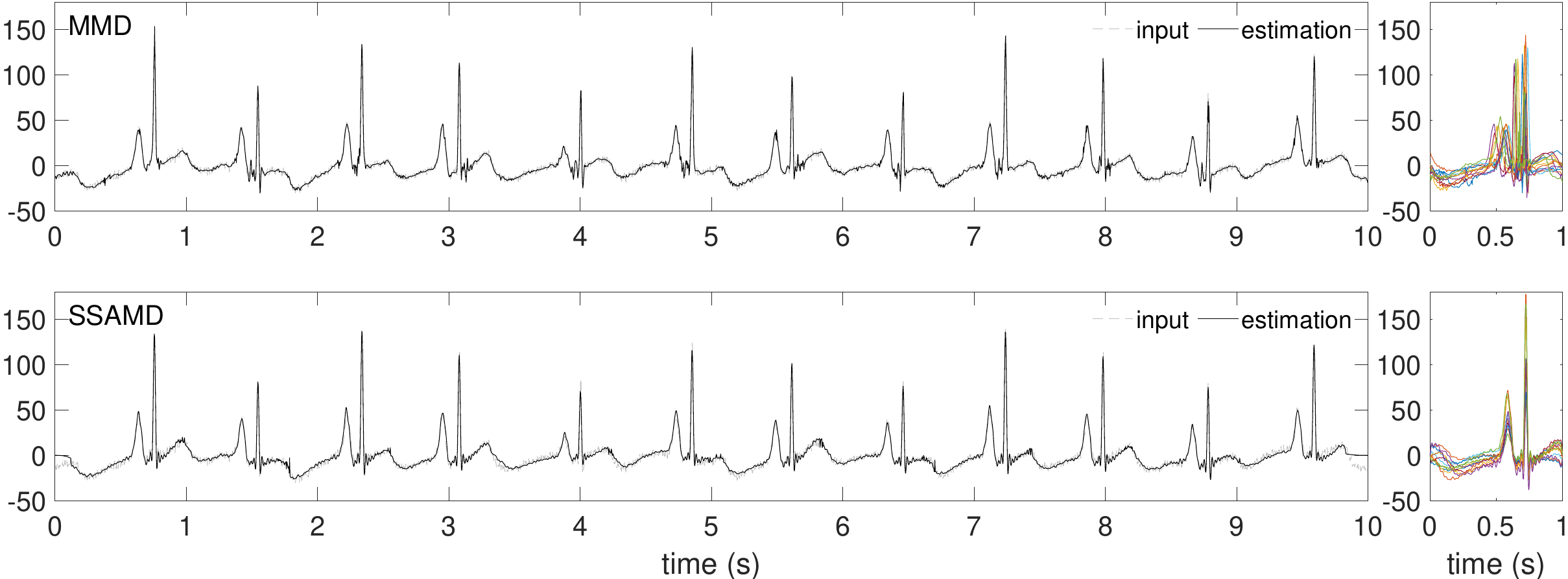}
\end{center} 
\caption{\textbf{Signal from the MIT-BIH Arrhythmia Database. Record 222m, channel 1.} Top row: input signal. Second row: LR results and its estimated waveform. Third row: SWM-fixed-Alg results and its estimated waveform. Fourth row: MMD results and its estimated waveforms. Fifth row: SSAMD results and its estimated waveforms.} \label{fig:04} 
\end{figure}

\begin{figure}[h!] 
\begin{center} \includegraphics[width=\columnwidth]{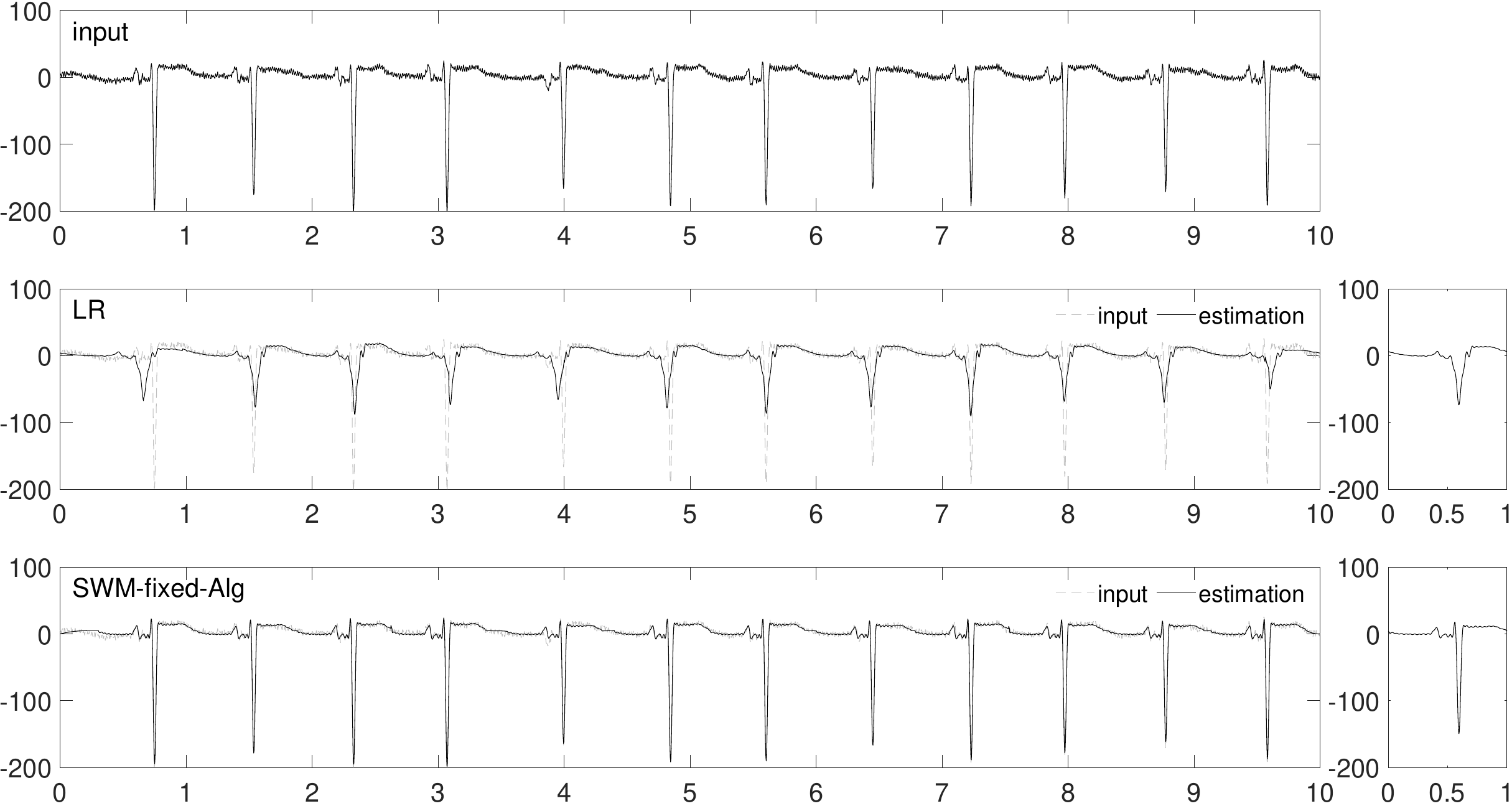} 

\vspace{2.5mm}
\includegraphics[width=\columnwidth]{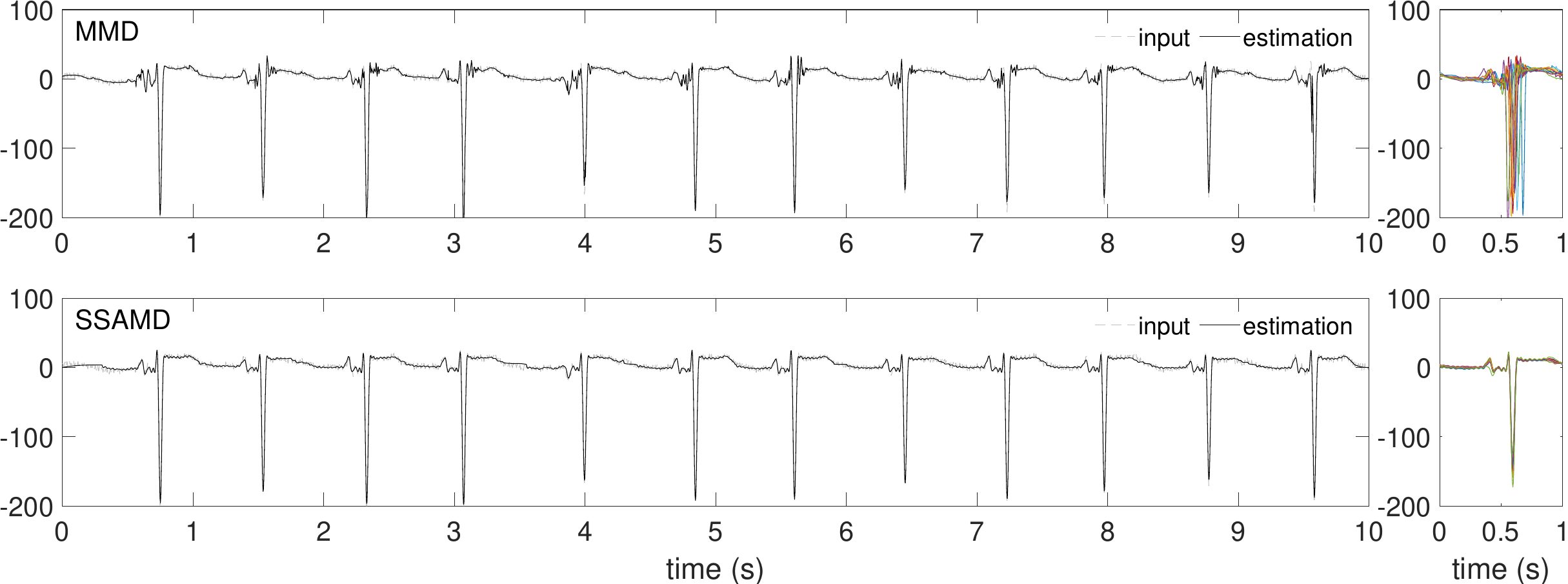}
\end{center} 
\caption{\textbf{Signal from the MIT-BIH Arrhythmia Database. Record 222m, channel 2.} Top row: input signal. Second row: LR results and its estimated waveform. Third row: SWM-fixed-Alg results and its estimated waveform. Fourth row: MMD results and its estimated waveforms. Fifth row: SSAMD results and its estimated waveforms.} \label{fig:05} 
\end{figure}

We present the results on two real ECG signals from the MIT-BIH Arrhythmia Database \cite{moody2001impact}. As before, we compare our results with those of the ANH model \emph{via} LR. Results are shown on Figs. \ref{fig:04} and \ref{fig:05}. For both LR and the new SWM-fixed-Alg, we can estimate one single waveform that repeats each cycle. On the other hand, SSAMD (with the SWM-variable model) allows for a cycle-to-cycle waveform estimation, better adjusting to the variations observed on real physiological signals (see fifth row). The results retrieved by MMD are displayed on the fourth row, and one can observe that the modes are noisier (see the fast oscillations around the QRS complexes), and the segmentation of the individual cycles is not as satisfactory.
Numerically, for this 10 s long signal sampled at 360 Hz, SWM-fixed-Alg takes 46.665 s, and SSAMD takes 48.54s, while MMD take 115.385 s.

In the signal from channel 2 (Fig. \ref{fig:05}), attention must be placed on the fifth cycle, just before the $4$th second. This cycle is different, as it can be observed on the negative deflection of the P wave. This difference can only be caught by MMD and SSAMD that allow for adjusting these slight variations, while MMD is noiser. Numerically, for this 10s long signal sampled at 360 Hz SWM-fixed-Alg takes 26.360 s, and SSAMD takes 27.376s, while MMD take 82.74 s.

\begin{figure}[h!] 
\begin{center} 
\includegraphics[width=\columnwidth]{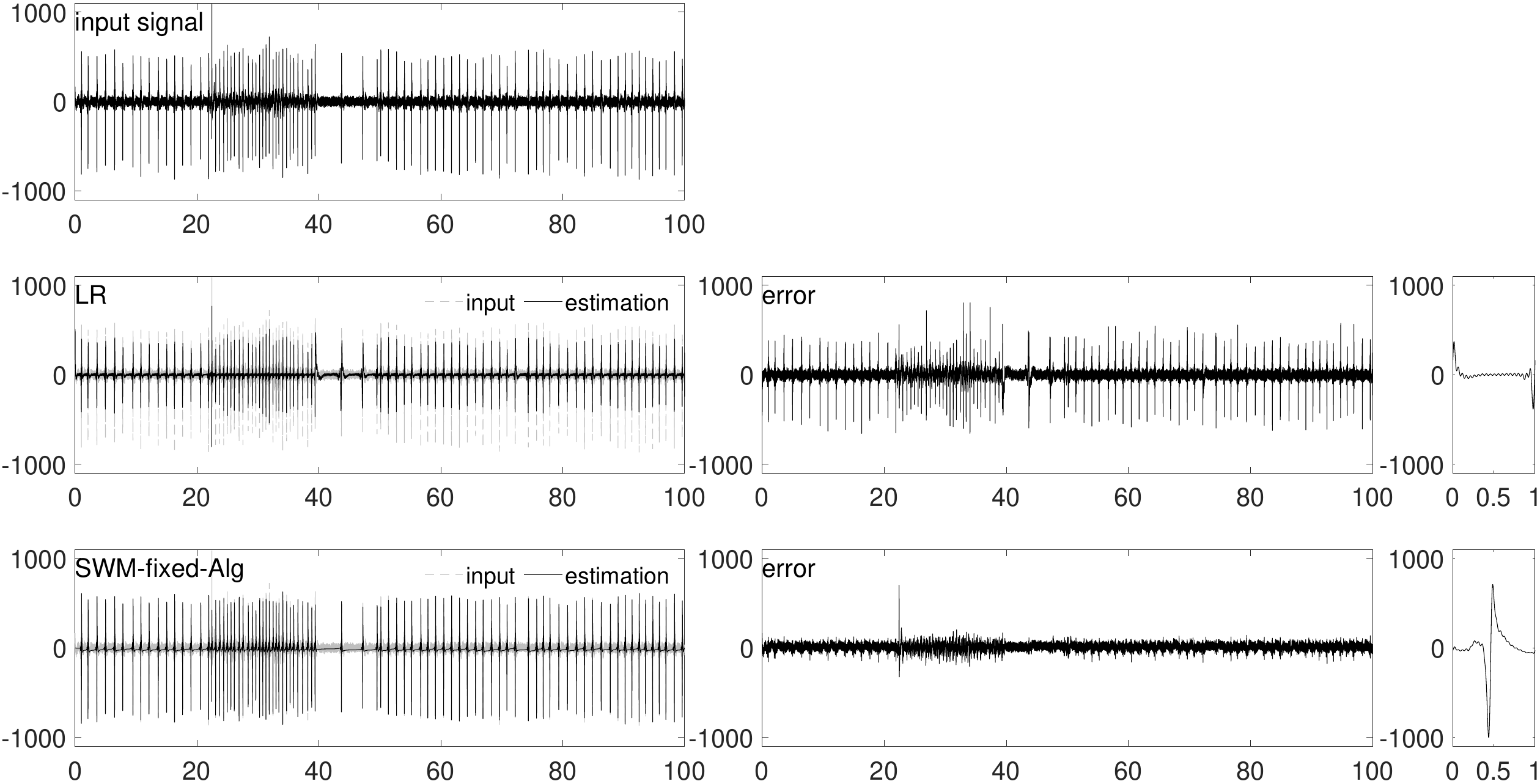}

\vspace{2.5mm}
\includegraphics[width=\columnwidth]{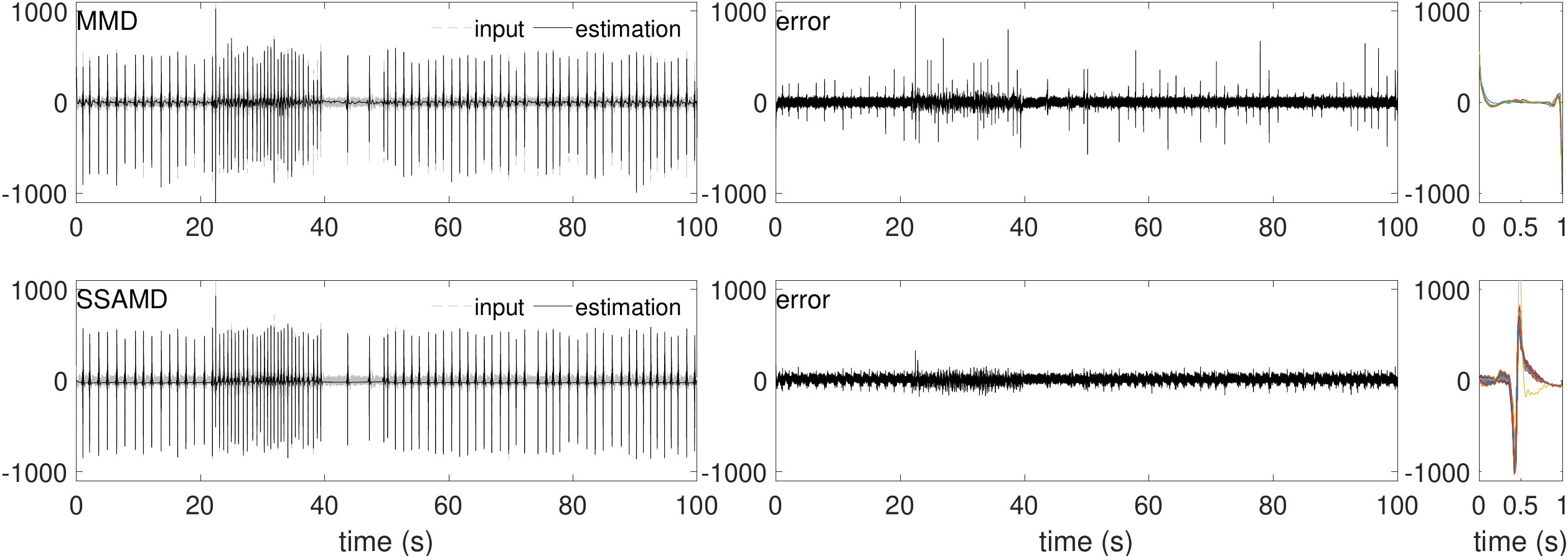}
\end{center} 
\caption{\textbf{ECG with Atrial fibrillation.} Top Row: the ECG signal recorded from a subject with persistent atrial fibrillation. Second row: LR results, its error, and its estimated waveform. Third row: SWM-fixed-Alg results, its error, and its estimated waveform. Fourth row: MMD results, its error, and its estimated waveforms. Fifth row: SSAMD results, its error, and its estimated waveforms.} \label{fig:AP} 
\end{figure}

Last, we consider a challenging ECG segment recorded from a subject with persistent atrial fibrillation. The results are shown in Fig. \ref{fig:AP}. We estimated the initial central time instants by detecting the QRS complexes, and also used these points to construct a rough estimate of the phase. The segmentations achieved by LR (under the ANH model) and MMD do not present the waveforms with the QRS complexes on the central positions, while SWM-fixed-Alg and SSAMD do. The errors are evidence of the superiority of SSAMD, where the QRS induced spikes are almost absent from them in the SWM-variable model (fifth row), while they remain prominent for LR and MMD.

In summary, these real ECG examples illustrate the denoising capabilities of SSAMD, and its capability of segmenting the signal into individual cycles. On the other hand, MMD offers noisier modes (again see the fast oscillation around the QRS complexes), and the segmentation into individual cycles presents some problems.

\subsection{Simulated Fetal ECG Extraction} 
\begin{figure}[h!] 
\begin{center} \includegraphics[width=\columnwidth]{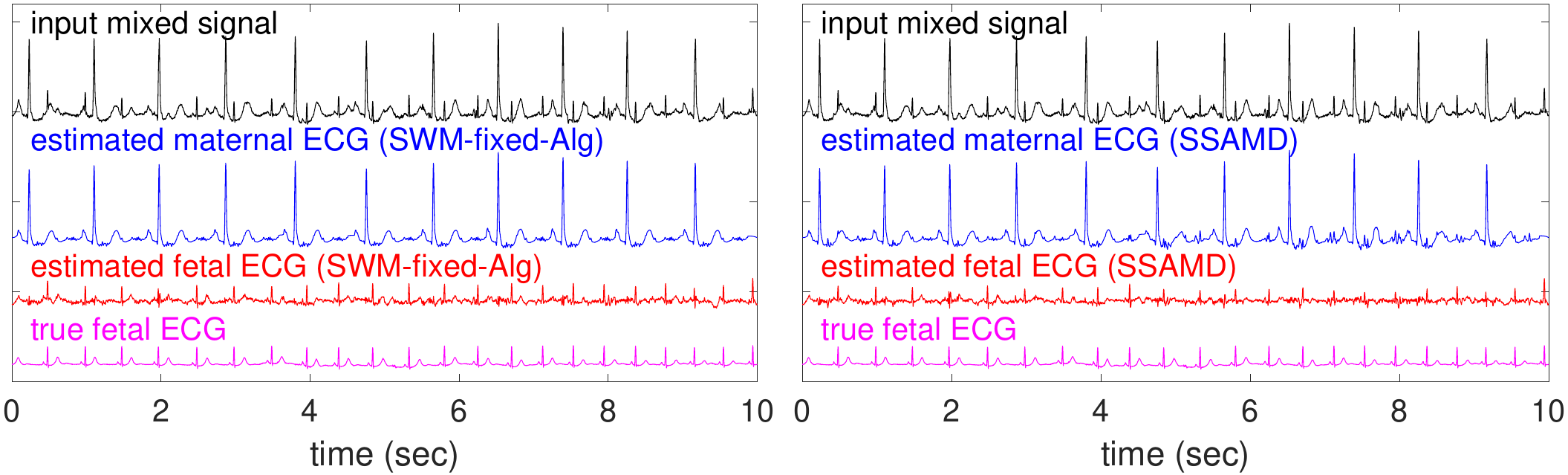} 
\end{center} \caption{\textbf{Simulated fetal ECG extraction.} Simulated maternal and fetal ECG mixture by taking signals from the MIT-BIH Arrhythmia Database. Left: results of SWM-fixed-Alg. Right: results of SSAMD.} \label{fig:sim_fetal} 
\end{figure}

Here we present a simulated example illustrating fetal ECG extraction. We simulated a mix of maternal and fetal ECG by taking a 10-second segment of Record 101m, channel 1, from the MIT-BIH Arrhythmia Database \cite{moody2001impact} as maternal ECG, and a 20.5-second segment of Record 106m, channel 1, from the MIT-BIH Arrhythmia Database as fetal ECG. The 20.5-second segment was resampled to match the 10-second segment, and it was multiplied by $0.2$. These simulated example is meant to illustrate the decomposition capabilities of our method, while having a ground truth to compare the results with. Fig. \ref{fig:sim_fetal} presents the results with both SWM-fixed-Alg and SSAMD algorithms. We took the mixed signal as input, and used the output of our respective models as estimations of the maternal ECG. The fetal ECG is estimated as the difference. In both cases (SWM-fixed and SWM-variable models) the estimated fetal ECG signals are of good quality when compared against the ground truth.

\subsection{Real Fetal ECG Extraction}

For this real examples, we deal with the fetal ECG extraction challenge, presenting results on three signals from the 2013 Physionet/CinC Challenge \cite{moody2001impact}. These are real abdominal recordings with both maternal and fetal components, and with no ground truth. We chose three signal with different maternal ECG morphology.

\begin{figure}[h!] 
\begin{center} 
\includegraphics[width=\columnwidth]{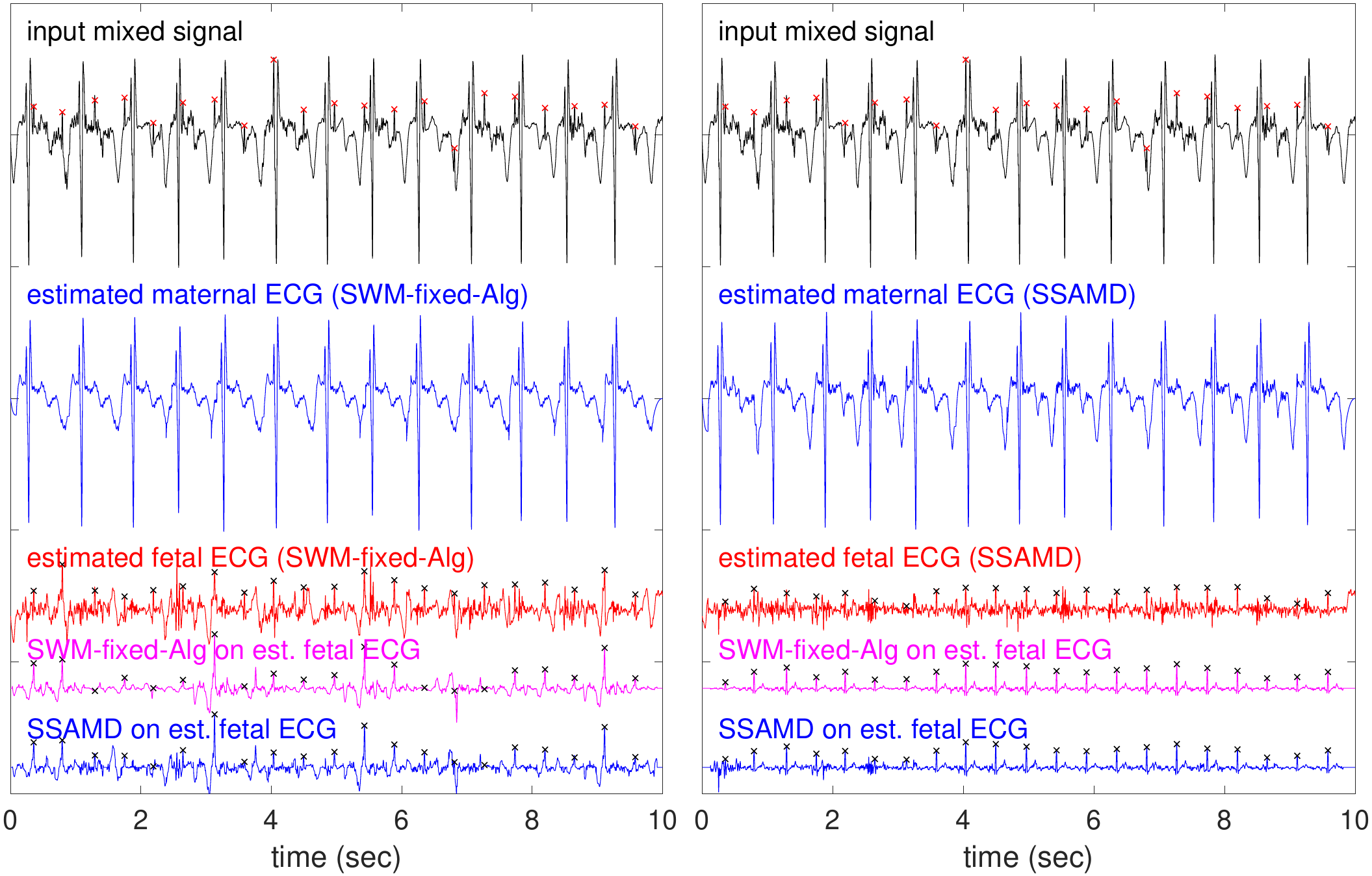} 
\end{center} \caption{\textbf{Real fetal ECG extraction.} Real abdominal ECG recording from the 2013 Physionet/CinC Challenge. Record a01, channel 02, first 10 seconds. Left: results of SWM-fixed-Alg. Right: results of SSAMD. In both cases, true annotated fetal R-peaks are marked with crosses.} \label{fig:real_fetal_01} 
\end{figure}

Figure \ref{fig:real_fetal_01} presents the results on the first ten seconds of channel 02 of record a01. The true annotated fetal R-peaks (taken from the database information) are marked with crosses. The left column shows the SWM-fixed-Alg results, and the right column does it so for SSAMD. We see that both estimated maternal ECG (in blue color) are acceptable. Denote the difference of the input trans-abdominal ECG and the estimated maternal ECG as $\tilde f_{\texttt{SWM-fixed-Alg}}$ and $\tilde f_{\texttt{SSAMD}}$, which are estimates of the fetal ECG (red color). Visually, $\tilde f_{\texttt{SSAMD}}$ performs better in the sense that fetal R peaks are more visible.

With the estimated fetal ECG, $\tilde f_{\texttt{SWM-fixed-Alg}}$ and $\tilde f_{\texttt{SSAMD}}$, we can apply the algorithm again to recover the fetal ECG. Although SSAMD does not need landmark points, one still can used them as initial estimations $\{a_i^{0}\}_{i=1}^I$ and $\{t_i^{0} \}_{i=1}^I$
if desired. To focus on demonstrating the capability of SWM-fiex-Alg and SSAMD on recovering fetal ECG, here we used the true annotated R-peaks as initial estimations. The results are presented on magenta color (SWM-fixed-Alg) and blue color (bottom line, for SSAMD). We can see that the denoised results of $\tilde f_{\texttt{SSAMD}}$ (right column) is better with both SWM-fixed-Alg and SSAMD. In this case, the magenta result seems to be better, mainly due to the presence of noise on the red signal, and the higher noise-robustness of SWM-fixed-Alg due to the fixed WSF. It should be noticed that the fetal ECG estimation not only provides with reliable R-peaks (useful to estimate the fetal heart rate) but also offers a good morphology estimation, where the fetal T-waves can be appreciated.

\begin{figure}[h!] 
\begin{center} 
\includegraphics[width=\columnwidth]{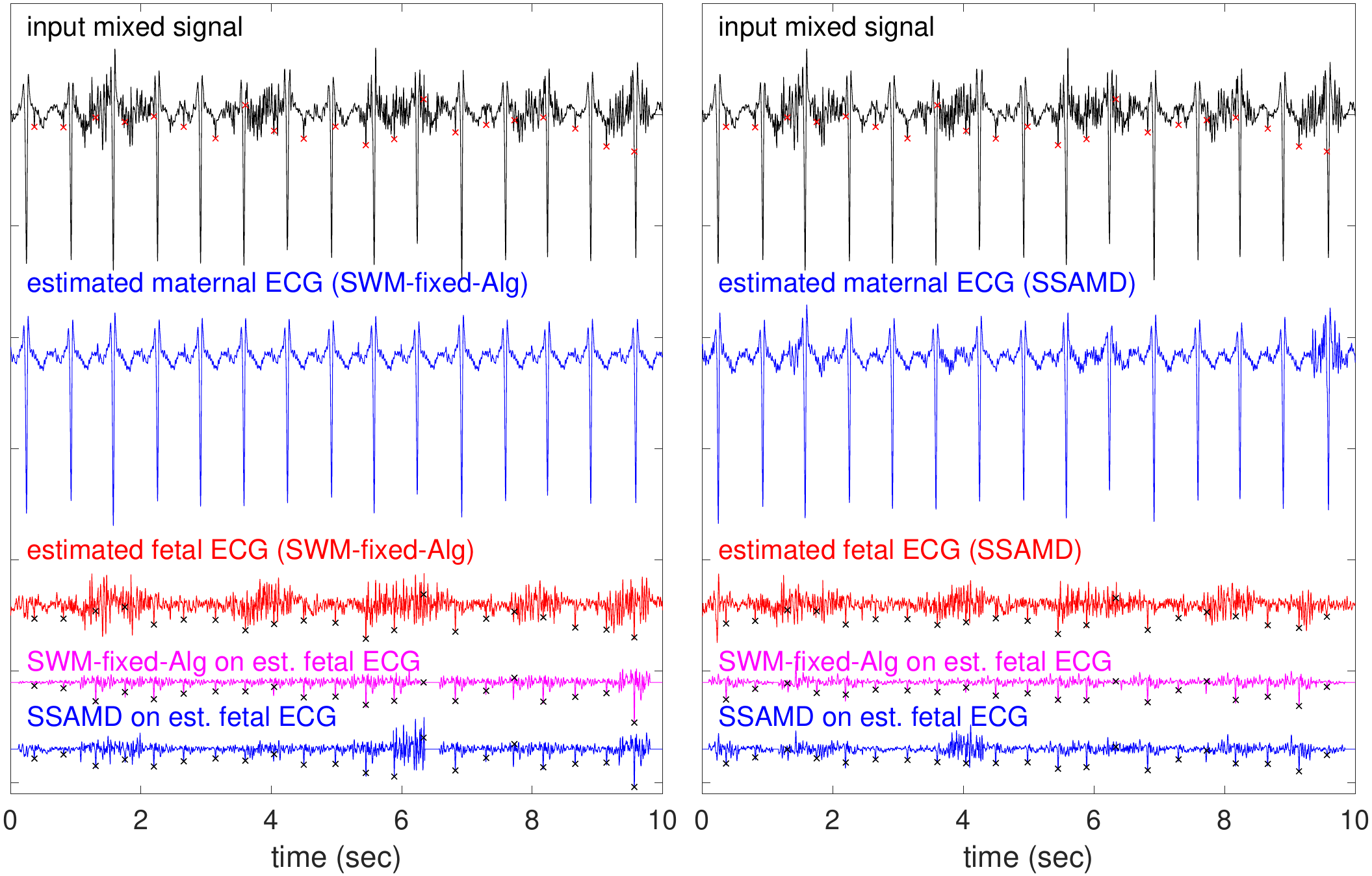}
\end{center} \caption{\textbf{Real fetal ECG extraction.} Real abdominal ECG recording from the 2013 Physionet/CinC Challenge. Record a07, channel 05, first 10 seconds. Left: results of SWM-fixed-Alg. Right: results of SSAMD. In both cases, true annotated fetal R-peaks are marked with crosses.} \label{fig:real_fetal_07} 
\end{figure}

A more challenging example can be found on Fig. \ref{fig:real_fetal_07}, where we analyzed the first ten seconds of the channel 05 of record a07. In this case, the fetal ECG is weaker with a higher noise and its R-peaks are negative deflections. The SSAMD results absorb some of the noise into de maternal ECG estimation, and this ends up being beneficial when estimating the fetal ECG as the difference (see red curves). Both SWM-fixed-Alg and SSAMD ran on $\tilde f_{\texttt{SSAMD}}$ retrieve acceptable results where the fetal R-peaks can be clearly identified.

\begin{figure}[h!] 
\begin{center} 
\includegraphics[width=\columnwidth]{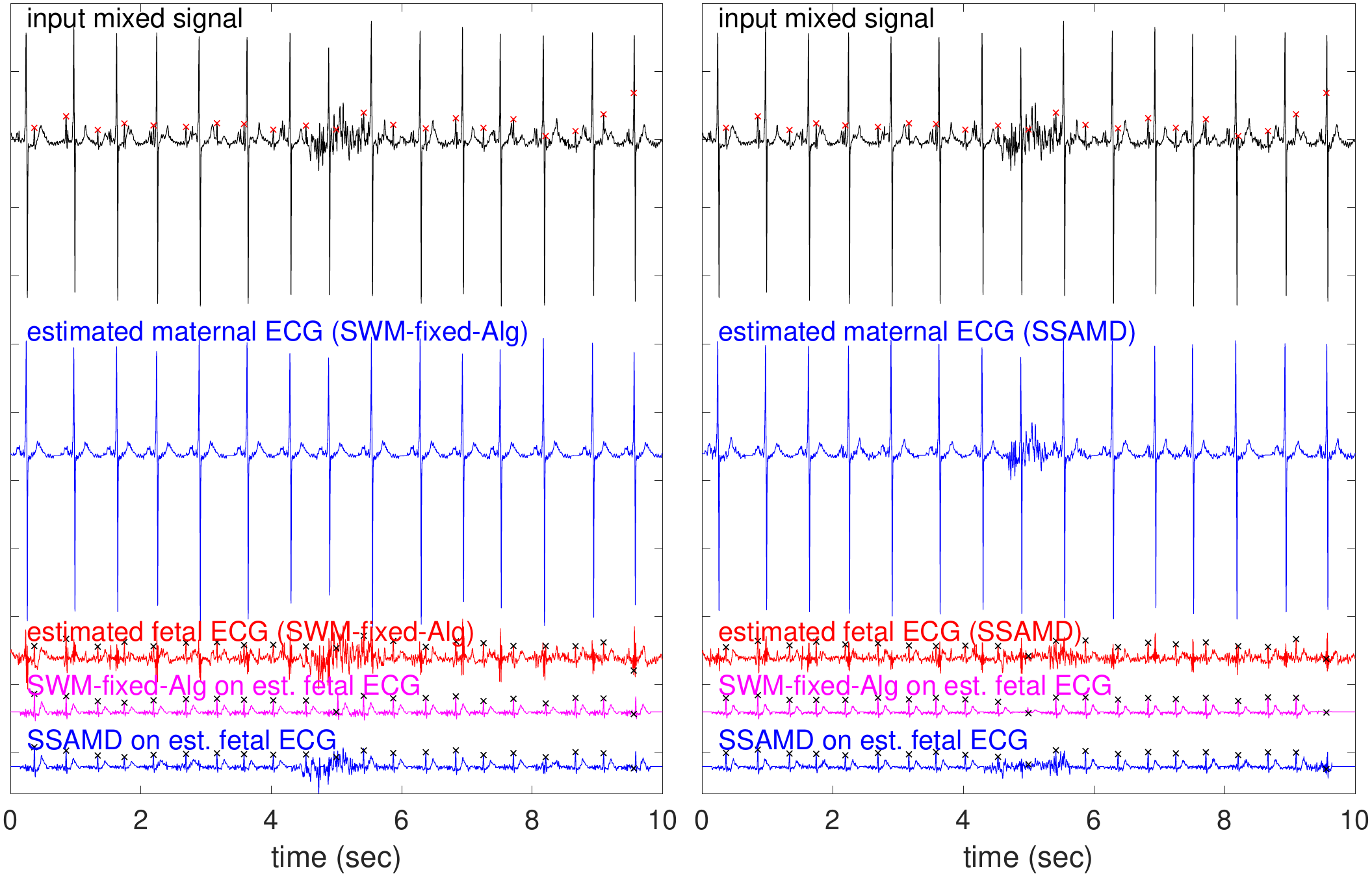} 
\end{center} \caption{\textbf{Real fetal ECG extraction.} Real abdominal ECG recording from the 2013 Physionet/CinC Challenge. Record a16, channel 02, first 10 seconds. Left: results of SWM-fixed-Alg. Right: results of SSAMD. In both cases, true annotated fetal R-peaks are marked with crosses.} \label{fig:real_fetal_16} 
\end{figure}

A final real example is shown on Fig. \ref{fig:real_fetal_16}, where the results for the first ten seconds of channel 05 of record a16 are presented. Here the maternal QRS complexes present deflections both positive and negative, and a burst of noise appears on the middle part of the signal. The post-processing of the first estimated fetal ECG components retrieves good results. Specifically, SWM-fixed-Alg applied to $\tilde f_{\texttt{SSAMD}}$ achieves the best results. The fetal T-waves can be clearly observed on this fetal ECG estimation (magenta curve, right column).

\section{Conclusions}

In this paper, we proposed a data-driven framework for modeling and estimating periodic signals with spiky WSFs. By representing individual WSF in the Fourier coefficient domain, the proposed approach performs dimension reduction to improve robustness to noise and artifacts. Starting from an initial fixed-WSF estimate, the waveform coefficients are refined through an optimization procedure based on an SVD-entropy regularization. This formulation promotes a compact representation of the time-varying WSF by encouraging morphological similarity across cycles while preserving physiologically meaningful variability.

We validated the proposed framework on both synthetic and real biomedical datasets. Experimental results demonstrated its effectiveness for denoising, signal decomposition, and waveform segmentation in challenging applications, including epileptic EEG, ECG, and single-channel fetal ECG extraction. Although motivated by biomedical signal processing, the proposed methodology is applicable to a broader class of periodic signals exhibiting localized, time-varying waveform morphology.

Several directions remain for future work. In particular, it would be of interest to establish a deeper theoretical connection between the proposed optimization framework and the underlying wave-shape manifold model, as well as to investigate adaptive manifold representations that further improve robustness under complex noise and artifact conditions.

\section*{Appendix}

Let $X_c=X-X_0$ and consider its singular value decomposition
\begin{equation}
X_c = U\Sigma W^\top,\qquad \Sigma=\mathrm{diag}(\sigma_1,\dots,\sigma_r).
\end{equation}

Define the normalized energy distribution
\begin{equation}
p_i = \frac{\sigma_i^2}{S}, \qquad S=\sum_{j}\sigma_j^2,
\end{equation}
and the SVD-entropy
\begin{equation}
H(X_c) = -\sum_{i} p_i \log p_i.
\end{equation}

We first compute the partial derivative of $H$ with respect to $\sigma_k$:
\begin{equation}
\frac{\partial H}{\partial \sigma_k}
= -\sum_i \frac{\partial p_i}{\partial \sigma_k}(1+\log p_i).
\end{equation}

Since
\begin{equation}
\frac{\partial p_i}{\partial \sigma_k}
= \frac{2\sigma_k}{S}(\delta_{ik}-p_i),
\end{equation}
with $\delta_{ik}$ the Kronecker delta, we obtain
\begin{equation}
\frac{\partial H}{\partial \sigma_k}
= -\frac{2\sigma_k}{S}
\left[(1+\log p_k)-\sum_i p_i(1+\log p_i)\right].
\end{equation}

Using the identities
\begin{equation}
\sum_i p_i = 1, \qquad \sum_i p_i\log p_i = -H,
\end{equation}
this expression simplifies to
\begin{equation}
\frac{\partial H}{\partial \sigma_k}
= -\frac{2\sigma_k}{S}(\log p_k + H).
\end{equation}

For spectral functions of the form $F(X_c)=\sum_i f(\sigma_i)$, the matrix gradient is given by
\begin{equation}
\nabla_{X_c} F = U\,\mathrm{diag}\!\left(\frac{\partial F}{\partial \sigma_i}\right)W^\top.
\end{equation}

Therefore,
\begin{equation}
\nabla_{X_c} H
= U\,\mathrm{diag}\!\left(-\frac{2\sigma_i}{S}(\log p_i + H)\right)W^\top.
\end{equation}

\bibliographystyle{plain}
\bibliography{biblio}

\end{document}